# Nuclear Waste Imaging and Spent Fuel Verification by Muon Tomography

## G. Jonkmans[a], V. N. P. Anghel[a], C. Jewett[a], M. Thompson[a]

[a]Atomic Energy of Canada Limited, Chalk River Labs, Chalk River, Canada, K0J 1J0

Corresponding Author: jonkmang@aecl.ca Tel: (+1) 613.584.3311 x43916, fax: (+1) 613.584.8198

## Abstract

This paper explores the use of cosmic ray muons to image the contents of shielded containers and detect high-Z special nuclear materials inside them. Cosmic ray muons are a naturally occurring form of radiation, are highly penetrating and exhibit large scattering angles on high Z materials. Specifically, we investigated how radiographic and tomographic techniques can be effective for non-invasive nuclear waste characterization and for nuclear material accountancy of spent fuel inside dry storage containers. We show that the tracking of individual muons, as they enter and exit a structure, can potentially improve the accuracy and availability of data on nuclear waste and the contents of Dry Storage Containers (DSC) used for spent fuel storage at CANDU® plants. This could be achieved in near real time, with the potential for unattended and remotely monitored operations. We show that the expected sensitivity, in the case of the DSC, exceeds the IAEA detection target for nuclear material accountancy.

## 1.     Introduction

Because of their unique ability to penetrate matter, cosmic ray muons are used to image the interiors of structures (Alvarez, *et al.*, 1970). Recently, a number of groups have extended the concept of muon

---

® Registered Trade Mark

radiography to the tracking of individual muons as they enter and exit a structure (Borozdin, *et al.,* 2003; Österlund *et al.*, 2006; Presente *et al.*, 2009; Gnanvo *et al.*, 2011). Most current efforts have been geared toward demonstrating the potential for muon tomography to detect the smuggling of Special Nuclear Materials (SNM) in cargo. This paper explores the application of muon tomography to non-invasive characterization of legacy nuclear waste containers and for nuclear material accountancy of spent fuel inside Dry Storage Containers (DSC) used to store spent CANDU fuel.

The paper reviews the concept of muon tomography and outlines how it applies to two examples of non-security related applications. This paper also presents the simulations and calculations conducted to demonstrate the efficacy of the technique and discuss the results and their implications.

## 2.    Muon Tomography

We refer to tomography as the reconstruction of the three-dimensional non-uniform matter distribution in a volume of interest from measurements made outside this volume. This differs from radiography in which only a 2-dimensional density map is obtained. In muon tomography we use the natural flux of highly penetrating cosmic-ray muons to produce images of the interiors of structures.

Muons, which are charged particles, much like electrons, are created by the interaction of cosmic radiation with the upper layer of the atmosphere.  Muons have the same electric charge as electrons, which make them easy to detect but, in contrast are much more massive ($m_\mu \approx 207 m_e$) and, on average, have a high momentum, which confers them with their strong penetrating ability. The vertical flux of energetic muons, above 1 GeV/c, at sea level is about 70 m$^{-2}$ s$^{-1}$ sr$^{-1}$ (or an integrated vertical flux $\approx 1$ cm$^{-2}$ min$^{-1}$) (Amsler *et al.*, 2008). The mean muon energy is 3-4 GeV, which is sufficient to allow them to penetrate several meters of rock before being stopped. The idea of using muons to image large structures is not new.  The first measurements date from 1955, when E.P. George measured the amount of overburden of the Guthega-Munyang Tunnel via the attenuation of cosmic rays (George 1955). Later, in 1970, Luis Alvarez dramatically demonstrated the use of cosmic-ray muons to look for hidden chambers in the Second Pyramid of Giza in Egypt (Alvarez, *et al.*, 1970). Muon radiographs are still in use today as a means to image structures of volcanoes in an attempt to predict volcanic eruptions (Nagamine 1995; Marteau *et al.*, 2011).

These previous measurements were all based on the attenuation of the muon flux. Recently, a number of groups have extended the concept of muon attenuation radiography to the tracking of individual muons as they enter and exit structures. In muon tomography, the incoming and outgoing directions are recorded for each muon, and the scattering location attributed to the Point of Closest Approach (PoCA) of the two tracks, thereby producing a 3D image of the scattering locations (Borozdin, *et al.,* 2003; Österlund *et al.*, 2006; Presente *et al.*, 2009; Gnanvo *et al.*, 2011). Multiple scattering of muons through material is mostly due to Coulomb scattering from nuclei. The distribution of the scattering angles of muons which passed through a target with charge number $Z$ is approximately Gaussian with a mean of zero and a width given by (Amsler *et al.*, 2008):

$$\sigma_\theta = \frac{13.6 MeV}{\beta c p} \sqrt{x/X_0} \left[1 + 0.038 \ln(x/X_0)\right] \qquad (1)$$

where $\beta c$ and $p$ are the muon velocity and momentum, and $x/X_0$ is the material thickness in units of radiation length. One can use Dahl's expression (Amsler *et al.*, 2008) for $X_0$ to further exhibit the $Z$ dependence of the width of the scattering angle distribution:

$$X_0 = \frac{716.4 \,\text{g cm}^{-2} A}{Z(Z+1)\ln(287/\sqrt{Z})}, \qquad (2)$$

with $A$ the atomic mass. Equation (1) is accurate to within 11% for $10^{-3} < x/X_0 < 100$, for all $Z$ and for $\beta c = 1$. Hence, the spread of scattering angles is larger for materials with high atomic numbers and small radiation lengths, and therefore the RMS of the deflection angle of scattered muons provides a means of discriminating between low, medium and high-Z materials.

The principle of a detector system for muon tomography is shown on Figure 1. Position sensitive detectors are placed above and below the structure to be scanned (i.e. container). A minimum of two such detectors is required for each top and bottom plane so that the muon incident and exit directions can be determined. With a simple algorithm, the image is reconstructed from the Point of Closest Approach (PoCA), i.e. the point in space that is the closest to both tracks.

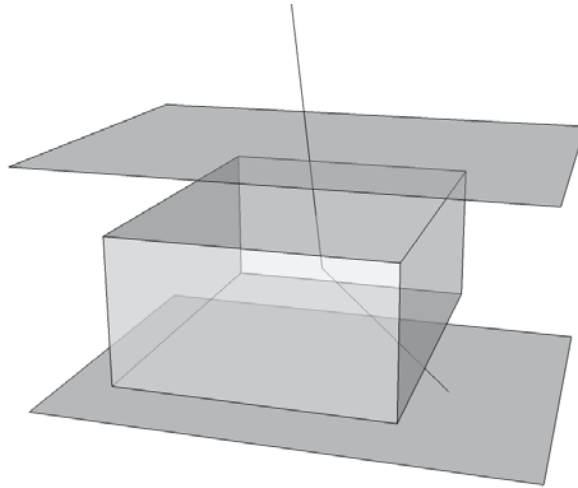

**Figure 1   Concept of muon tomography.**

Several methods have been proposed to discriminate further between low-Z and high-Z materials (Österlund *et al.*, 2006; Schultz, *et al.,* 2003; Schultz *et al.,* 2004; W. Priedhorsky, *et al.*, 2003; Green, *et al.*, 2006) and a number of groups have carried out developments in muon tomography. Some have focused on a system for the purpose of detecting contraband fissile material in cargo (Schultz, *et al.,* 2003; Schultz *et al.,* 2004; Borozdin, *et al.*, 2003) while others are developing a system for tomography of canisters of spent nuclear fuel (Donnard 2004; Flodin 2005; Gustafsson 2005; Jonkmans *et al.*, 2010).

## 3.      Non-security applications of muon tomography

### 3.1   Nuclear waste characterization

AECL's Nuclear Laboratories maintain a radioactive waste management program to protect public health and the environment and to ensure worker safety. This is partly implemented by a program to segregate and characterize nuclear waste. Generally, radioactive waste exhibits a wide spectrum of physical, chemical and radiological characteristics, which demands a complex and tailored waste characterization program.  Although no single methodology can be adopted to treat such varied waste, strategies are beginning to emerge to develop guidance in radioactive waste characterization to facilitate long-term storage or disposal (see for example IAEA 2007a and references therein).  Current practice calls for establishing a process for characterizing radioactive waste during its entire life cycle, which means that the type of waste and waste form, the evolution of the waste with time, and the measurements to characterize the waste are usually identified upstream of segregation and disposal. Hence this is typically the case for *new wastes* as defined as those that are generated with a traceable and robust characterization program in place.

In contrast, *historical wastes* lack a comprehensive traceable characterization program; hence it makes the process for long-term storage or disposal more difficult.  In short, this type of waste has a non-traceable waste stream, has an incomplete history and incomplete or improper characterization.  In several cases this results in a large amount of uncertainty as to the nature of the waste: origin, traceability and conditioning.  This is often the case with legacy nuclear waste containers[1]. Although there is a large amount of variability of waste types and origin, one can identify two components that a robust characterization program must address:

- waste complexity, which refers to the level of difficulty to develop a list of properties (*fingerprint*) to characterize the waste;

- waste stability, which refers to whether or not these properties remain relatively constant with time.

Those components refer not only to the waste content but also to the waste package. The key elements of the characterization program involve fingerprinting the waste's radioactivity, chemical, physical, mechanical, thermal and biological properties and how they evolve through time.

Challenges and issues for nuclear waste characterization are covered at great length in reference (IAEA 2007a). A significant subset of these is summarized below:

- Historical waste.  Non-traceable waste stream poses a challenge for characterization. Different types of waste can be distinguished: tanks with liquids, fabrication facilities to be decontaminated before decommissioning, interim waste storage sites, etc.

- Some waste form may be difficult and/or impossible to measure and characterize (i.e. encapsulated alpha/beta emitters, heavily shielded waste).

---

[1] It should be emphasized that *new* and *historical*, in this context, do not refer necessarily to the age of the waste, i.e. nuclear waste generated today, without a robust characterization program, is historical waste according to  this IAEA classification.

- Direct measurements, i.e. destructive assay, are not possible in many cases and Non-Destructive Assay (NDA) techniques are required, which often do not provide conclusive characterization.

- Homogeneity of the waste needs characterization (i.e. sludge in tanks, in-homogeneities in cemented waste, etc.).

- Condition of the waste and waste package: breach of containment, corrosion, voids, etc.

The use of muon tomography to assist in the characterization of nuclear waste is explored further in sections 4 and 5.

## 3.2    Non-proliferation and safeguards

Spent nuclear fuel is a form of nuclear waste with the added complexity of security and proliferation issues, and, as such, it is often placed under IAEA Safeguards. *IAEA Safeguards are measures through which the IAEA seeks to verify that nuclear material is not diverted from peaceful uses* (IAEA 2007b). *Traditional Measures* target the detection of diversion of nuclear material. It is the historical cornerstone of safeguards and focuses on material verification activities and containment and surveillance (C/S) at facilities where the states have declared the presence of nuclear material subject to safeguards. IAEA Safeguards' item accountancy objectives are defined by the *timely* detection of a *significant amount* of nuclear material (IAEA 1999). The detection targets are summarized in Table 1.

Clearly the use of muon tomography for verifying and accounting for spent fuel in dry storage is an attractive means by which to keep track of spent fuel inventories in a non-invasive, remote and unattended manner. In Canada, spent nuclear fuel is stored in large pools (fuel bays or wet storage) for a nominal period of 10 years to allow for sufficient radioactive cooling. Above ground dry storage is then used for the storage of used fuel. This method, considered an interim measure until an approach for long-term management of used fuel is implemented, was pioneered by AECL and Ontario Power Generation (OPG) (Stevens-Guille and Pare 1994). These above ground facilities are essentially canisters that are used as containment to protect workers and the environment from the remaining radiation and to safeguard the spent fuel. Consequently, site storage systems must satisfy stringent safety, security and environmental requirements. The prevalent methods of dry storage are: Dry Storage Containers (DSCs) and monolithic air-cooled concrete structures for dry storage (MACSTOR – Modular Air Cooled STORage) (Dowdeswell 2003; NWMO 2003).

DSCs, such as the one shown on Figure 2, are used throughout OPG facilities. The DSC is a reinforced high-density concrete container with an inner steel liner and an outer steel shell. The DSC can store up to 384 used fuel bundles in four standard storage modules of 96 bundles each. The dimensions of the DSC are 2.12 x 2.15 $m^2$ by 3.55 m high. Two stainless steel (SS) tubes are imbedded diagonally across the lid to provide for the installation of IAEA seals. Similarly, two U-shaped SS tubes are imbedded in the walls of the DSC and are used as inspection tubes by the IAEA.

**Table 1 IAEA detection targets for safeguards.**

| Item | Quantity [kg] | Encapsulation | Time to detect diversion |
|------|---------------|---------------|--------------------------|
| $^{235}$U | 75 | <20% enriched in NU or DU | 12 months |
| $^{235}$U | 25 | ≥20% in HEU | 1 month |
| Pu | 8 | All isotopes fresh fuel | 1 month |
| Pu | 8 | All isotopes spent (irradiated) fuel | 3 months |
| $^{233}$U | 8 | Fresh fuel | 1 month |
| $^{233}$U | 8 | Spent irradiated fuel | 3 months |

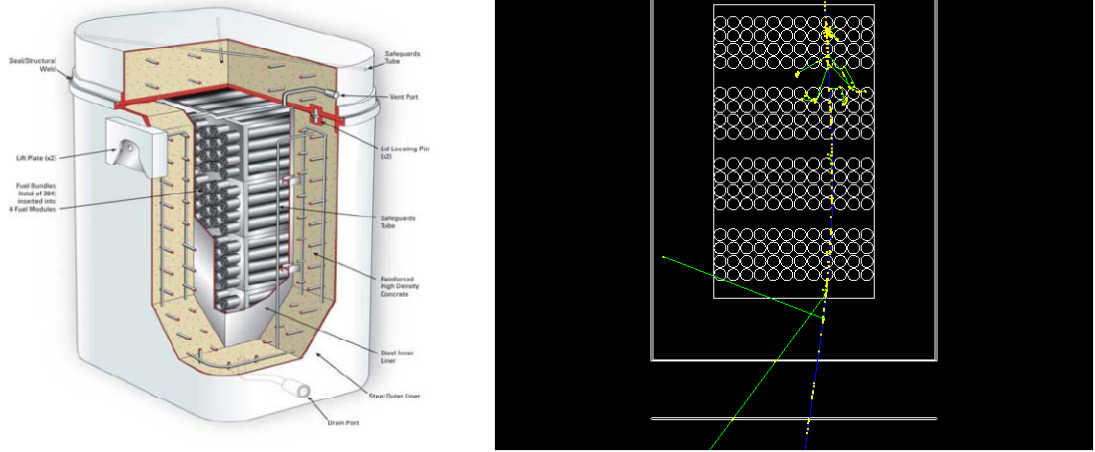

**Figure 2**   Cutaway drawing of a Dry Storage Container (DSC) (NWMO 2003) on the left and GEANT4 simulation showing a muon track and associated photons on the right.

## 4.      Simulations and image reconstruction

Evaluations of the efficiency of muon tomography to identify nuclear materials in containers were performed with Monte Carlo (stochastic) calculations.  The main advantage of performing Monte Carlo calculations is that they allow an event-by-event treatment, so that each muon is tracked independently, as it would be in a muon tomography system.  Evidently, simulating the scattering of muon tracks inside containers does not in itself characterize the nuclear material detection performance of a muon tomography system.  This must be addressed by image reconstruction techniques and/or statistical analysis of muon tracks.  This section describes the simulations, image reconstruction methods and derived detection performance for a selected number of container scenarios.

### 4.1   Simulations

Muon scattering in containers was simulated with the GEANT4 software (Agostinelli *et. al.*, 2003; GEANT4 2007a; GEANT4 2007b). GEANT4 is a simulation toolkit designed for modeling a broad range of particle processes and their interaction with matter. It is used in a variety of applications, including High Energy Physics (HEP), nuclear physics experiments, astrophysics, space science and medical physics. Its distinguishing feature is that it allows users control over each typical domain of a simulation: geometry, particles, run, events, tracking, detector response, physics processes and transparent access to cross-sections.

For the purpose of this work the detector geometry is generic and is made of two thin detector planes (trackers): one set at the top of the container and the other at the bottom. Each tracker plane is made of 1 cm thick argon at standard temperature and pressure (STP). The trackers are idealized detectors in that they register the exact muon tracking information: muon track position, direction and momentum as they cross the detection plane. Also, the trackers are made insensitive to other particles, in particular they are insensitive to track multiplicities (showers of particles) in the bottom tracker, sometimes created by high-energy muons when hitting the container. The geometry of the container volume is variable and depends on the case under study.

Only the muon component of cosmic rays has been simulated. The event generator produces muons with realistic energy and angular distributions. We used an angular distribution of muons, which is peaked at the zenith and falls off as $\cos^2 \theta$, where $\theta$ is the angle from the vertical. The muon energy distribution was sampled from the $\mu^+$ distribution from the Tsukuba 95 measurement, BESS spectrometer data, Table 1 of reference (Motoki *et al.*, 2003), with a cut-off momentum of 0.576 GeV/c. All the most significant particle types were considered (*declared*) in the simulations (gammas, leptons, n, p and low mass mesons), but only the most significant processes were considered (for example, muon capture and muon decay in-flight were neglected). Specifically for muons, multiple scattering, ionisation, bremsstrahlung and pair production were declared. Also, a threshold, below which no secondary particles will be generated, is defined during this initialization phase. This production cut parameter for the simulation was generalized for all particles as a range cut of 1 mm, which is internally converted by GEANT4 into an equivalent energy for different materials. There are no other tracking cuts. All particles are tracked down to the range cut. This could be relaxed later to speed-up simulations. Muons are detected as they cross the detector planes: the coordinates of the track crossings and the momentum vector are recorded. Therefore, the detection system is an idealized representation and experimental effects of detector resolution, particle multiplicity and other effects have not been considered here.

## 4.2   Image reconstruction methods

There are several different methods of imaging high-Z material using muon tomography. We have so far arbitrarily chosen two imaging techniques: the Maximum Likelihood/Expected Maximization (ML/EM) tomographic reconstruction algorithm as described in references (Schultz *et al.,* 2007; L. Schultz, *et al.,* 2006) and a Scattering Density Estimation algorithm (SDE) developed by our group. Both methods exploit, in addition to the tracks' directions, the muons' momentum information. The ML/EM method was used with a slightly different approach in the scattering strength distribution of voxels.

### 4.2.1   Maximum Likelihood/Expected Maximization (ML/EM) tomographic reconstruction algorithm

The ML/EM tomographic reconstruction method, developed by the Los Alamos group and described in references (Schultz *et al.,* 2004) and (Schultz *et al.,* 2007), was the first method we implemented. The method uses as input the directions and momentum for each muon on the top and bottom surfaces of the volume of interest and assigns a scattering density $\lambda$ per unit length of material $H$:

$$\lambda = \frac{\sigma_\theta^2}{H}\left[\frac{\beta cp}{(\beta cp)_0}\right]^2, \tag{3}$$

for each voxel, where $\beta cp$ is the product of the muon speed and momentum, and $(\beta cp)_0$ the average speed times momentum for cosmic ray muons. The ML/EM algorithm considers the probability function, or log likelihood, of the scattering density function against the measured scattering density of each muon in voxels within the volume under scrutiny. We refer the reader to reference (Schultz *et al.,* 2004) for additional details. When applying this method, we have modified somewhat the definition of the scattering segments in the scattering strength. This work will be described in future publications.

### 4.2.2 Scattering Density Estimation (SDE) algorithm

The Scattering Density Estimation (SDE) algorithm (Anghel *et al.*, 2011) first proceeds by calculating the Point of Closest Approach (PoCA) to approximate the point of scattering of each detected muon through container volumes. One assumes that the scattering occurred due to a single scattering event, and locates that point by extrapolating the incident ($\boldsymbol{\mu}_1$) and outgoing muon trajectory ($\boldsymbol{\mu}_2$) to their PoCA. Hence, each muon track is defined by $\mathbf{r}$ and $\mathbf{q}$ where $\mathbf{r}$ is the coordinate of the muon trajectory crossing a tracker plane and $\mathbf{q}$ the momentum vector of the trajectory. The PoCAs on each trajectory are then easily computed from the line equations, $\mathbf{l}_1$ and $\mathbf{l}_2$,:

$$\mathbf{l}_1(s) = \mathbf{r}_1 + s\mathbf{q}_1 \text{ and } \mathbf{l}_2(t) = \mathbf{r}_2 + t\mathbf{q}_2, \tag{4}$$

such that $\left| \mathbf{l}_1 - \mathbf{l}_2 \right|$ is a minimum and where $s$ and $t$ are scalar parameters. The point of scattering is given as the midpoint between the two PoCAs. Plotting these points in itself does not usually reveal high-Z materials, since muons will also scatter in the surroundings. Normally additional cuts are needed on the scattering angle at the expense of reducing the data sample size.

Multiple scattering of muons in surrounding materials and measurement inaccuracies lead to errors in the PoCA calculations, which translate to blurring and localized false hot spots in the image. It was observed empirically that high scattering density points truly associated with the presence of high-Z materials are less affected by these inaccuracies. To lessen these effects, we used a *pitchfork* method which consists of adding 4 satellite tracks to the incoming muon trajectory and 4 satellite tracks to the outgoing muon trajectory. This is illustrated in Figure 3. The diagram shows the 5 vectors involved. The central vector (solid line) is the measured incoming or outgoing muon trajectory. The 4 outer prong trajectories (dashed lines) are defined by adding and subtracting a randomly-generated set of two perpendicular vectors (dotted lines) to and from the measured muon trajectory. The dotted vectors are themselves in a plane perpendicular to the muon trajectory. The magnitude of the vectors is made to coincide with the system resolution, taken to be 2 mrad in the current demonstration. The Scattering Density Estimation (SDE) algorithm uses a different definition of the scattering density but is close to the form of Equation (3):

$$\lambda_{voxel} = \frac{\left[\sum_i w_i \left|\hat{\boldsymbol{\mu}}_2^i - \hat{\boldsymbol{\mu}}_1^i\right| \cdot \left|\beta c \mathbf{q}_2^i\right|\right]^2 \sum_i w_i \left(\left|\mathbf{d}_i\right\|\beta c \mathbf{q}_2^i\right|\right)^2 \left|\hat{\boldsymbol{\mu}}_2^i - \hat{\boldsymbol{\mu}}_1^i\right| \cdot \left|\beta c \mathbf{q}_2^i\right|}{\sum_i w_i \left(\left|\mathbf{d}_i\right\|\beta c \mathbf{q}_2^i\right|\right)^3}, \qquad (5)$$

where the sums are taken over the 25 possible track pairs that belongs to the voxel identified by the PoCA. $\hat{\boldsymbol{\mu}}_1$ and $\hat{\boldsymbol{\mu}}_2$ are unit vectors along the incoming (outgoing) trajectories of the muon respectively, such that $\hat{\boldsymbol{\mu}}_2 - \hat{\boldsymbol{\mu}}_1$ is proportional to the scattering angle. $\left|\mathbf{d}_i\right|$ is the distance of closest approach between the trajectories. $\mathbf{q}_2$ is the momentum vector of the outgoing trajectory. $w_i$ is the weight for the pair, and is given by the product of the incoming and outgoing track weight (1/3 for the muon trajectories and 1/6 for the satellite trajectories). We note that the first summation in the numerator of Equation (5) takes the form of the change in momentum and the second summation takes the form of the change in angular momentum between the incoming and outgoing muon trajectories.

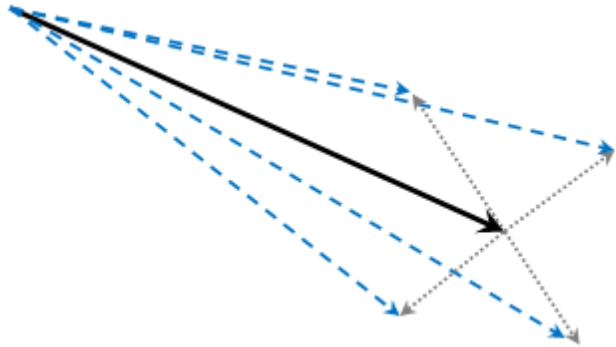

**Figure 3 Diagram of the 5 vectors involved in the *pitchfork* method. The central vector (solid line) is the measured incoming or outgoing muon trajectory. The 4 outer prong trajectories (dashed lines) are defined by adding and subtracting a randomly-generated set of two perpendicular vectors (dotted lines) to and from the measured muon trajectory.**

## 5.    Results and discussions

## 5.1   Waste characterization

The diversity of nuclear waste characteristics makes it unrealistic to look at all possible scenarios in which muon tomography can be used to characterize waste; however an example was chosen to illustrate the potential of this technology.  This example is tailored to the potential advantages of muon tomography: Non-Destructive Analysis (NDA) capability, sensitivity to atomic number, and imaging capability of content and containers.  Also, this particular example benefits from a high Z-contrast ratio, which makes it ideal for the development of imaging algorithm. The following case has been analyzed: fuel pencil (10 cm long, 1 cm diameter $UO_2$) encapsulated in a 55-gallon stainless steel drum. The drum is filled with paraffin wax. A 10 cm diameter void has also been added to simulate breach of containment.

The fuel pencil was positioned parallel to the central axis of the drum, but some distance away from the centre. One day's worth of cosmic ray muons was originally generated or about 400,000 events. Figure 4 shows a 2D projection of the canister using the ML/EM reconstruction algorithm (60 iterations) and the SDE algorithm. The pin location is prominent using both methods.

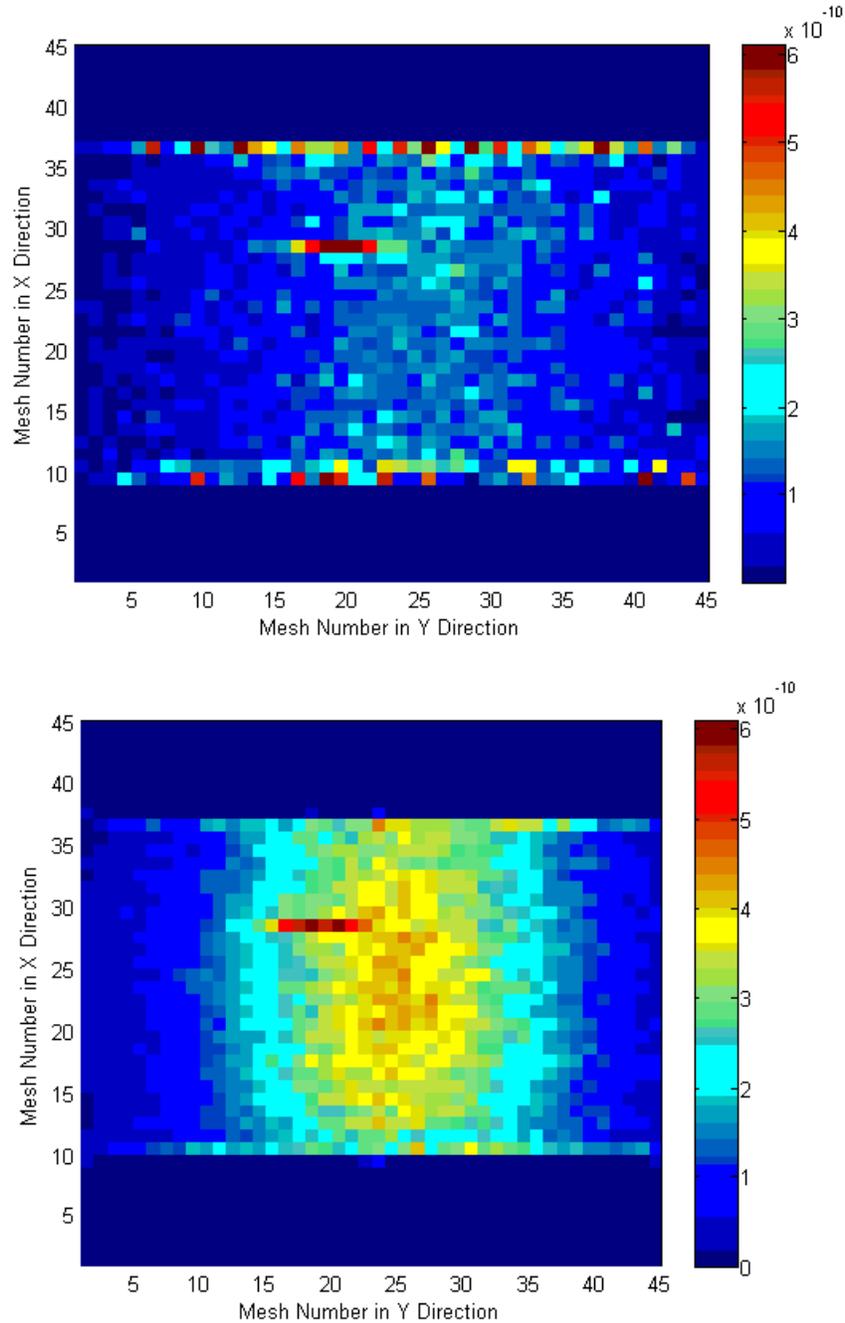

**Figure 4   Top: ML/EM reconstructed image of a UO₂ fuel pin inside a stainless steel drum on its side. Bottom: reconstructed image using the SDE algorithm. The mesh size in this image reconstruction is 2 cm. Scale is scattering density per voxel (arbitrary units).**

## 5.2          Spent fuel monitoring results

The problem of monitoring spent fuel in DSCs, using muon tomography, is vastly different than that of waste characterization and contraband detection. First, the container and content under scrutiny are assumed to be well known and one looks for deviations from a known configuration. Secondly, the safeguards requirements of timely detection and significant quantity (see Table 1) are generally less stringent than those for contraband detection, and a significant amount of data can be acquired to draw conclusions. Thirdly, the assumption of a single scattering point in PoCA-based methods can no longer be used owing to the large amount of spent fuel gathered in one large volume. Therefore the image reconstruction techniques devised so far fail to produce meaningful images.

To evaluate the ability of muon tomography to safeguard spent fuel, we have simulated a single type of spent fuel container modelled after the DSC with some simplifications. Figure 2 right shows a cutaway view of the simulated geometry. The model contains the same number of CANDU fuel bundle as a full DSC: 384 used fuel bundles in four standard storage modules of 96 bundles each. Each row is 2 bundles in length. The walls of the DSC where modelled as a concrete (density 2.3 g/cm$^3$) box of dimensions 2.12 m x 2.15 m by 3.55 m tall with a wall thickness of 52 cm. The concrete is surrounded by a thin, 1.3 cm, carbon steel shell (density 7.87 g/cm$^3$).

Figure 2 also illustrates a typical muon event. The extent of the continuous scattering, and the large number of multiple interactions, visible through secondary particles, of the muon with the fuel and the concrete is apparent in the figure. Muons typically lose about 1 GeV while going through the DSC, so a large fraction of the muons is stopped before reaching the bottom detector plane. As is evident from the figure, the continuous scattering of the muon, coupled with the homogeneity of the materials, does not permit one to define a single scattering point. Previous trials showing the PoCAs' distribution for two stringent angular cuts of 5° and 20° have shown that fuel bundles cannot be distinguished in the picture.

Although the PoCA-based methods described here are inapplicable to the reconstruction of the image of large amounts of dense material, muon radiography is expected to show density variations. Figure 5 shows a radiograph of a DSC with 4 columns of fuel missing. Our earlier attempts at producing muon radiographs from flux measurements of outgoing muons did not produce flux profiles from which an accurate density plot could be produced. This was attributed to the $\cos^2 \theta$ angular distribution of cosmic ray muons, and hence, the large differences in scattering strength for muons recorded in similar areas of the bottom detector plane. A flux profile was then built by counting only through-going muons. The accounting was performed by registering only muons which passed through overlapping vertical regions: a muon was registered if a hit was recorded in a top region (cell) and a bottom cell directly below it. Effectively, this method amounts to imaging the effect of the container on a parallel beam of muons. Figure 5 shows the radiograph obtained for the case with 5 cm x 5 cm cells. The green lines indicate the actual locations of the fuel bundles. The central region, where fuel bundles are missing, shows a much higher muon flux and is clearly discernable along with the concrete walls and the fuel bundles. This radiograph was produced with 12 million muons or 1 day equivalent. About 48,000 muons passed the cuts described above or 0.4 % of the total number of generated cosmic muons.

To further quantify the ability of muon radiography to meet the IAEA Safeguards' item accountancy objectives, a DSC container with 2 contiguous fuel bundles removed was simulated. In this simplified model, each fuel bundle is a simplified full cylinder, with the density of $UO_2$ adjusted so that the total bundle mass is 20 kg. Assuming that spent bundles contain about 0.35 wt % $^{235}$U and the same amount of $^{239}$Pu, this represents a total content per bundle of about 0.07 kg of $^{235}$U and 0.07 kg of $^{239}$Pu. Muon radiographs (parallel beam method) were obtained from a simulation of 7 days' worth of muons (517 million muons) of which 342,000 remained after cuts (0.07 %). In this analysis, an 11.3 cm cell size along the x-axis and an 8.4 cm cell size along the z-axis were used. The cell size was chosen such that an integral number of cells covered the fuel bundles.

Figure 6 shows the muon counts projected on one axis of the radiograph. The high point, at position $x = 17$, shows $29817 \pm 173$ counts, which is $5\sigma$ away from the mean, indicated by the orange line on Figure 6. Therefore, within 7 days, muon flux measurements on a DSC could be sensitive to a loss of 2 bundles with a high degree of confidence. Although, the flux radiograph on its own is not able to separately identify actinide content, it is likely that this kind of measurement, compared with the expected attenuation of muons as a function of material density, will provide a way to perform non-invasive material accountancy of nuclear materials with a high degree of confidence that greatly exceeds the IAEA detection targets shown in Table 1. Note that the low points at the extremities of the graph in Figure 6 are explained by edge effects. Muons which scatter in edge bundles and in the concrete have a smaller probability of re-entry into the bundle column in which flux is being computed.

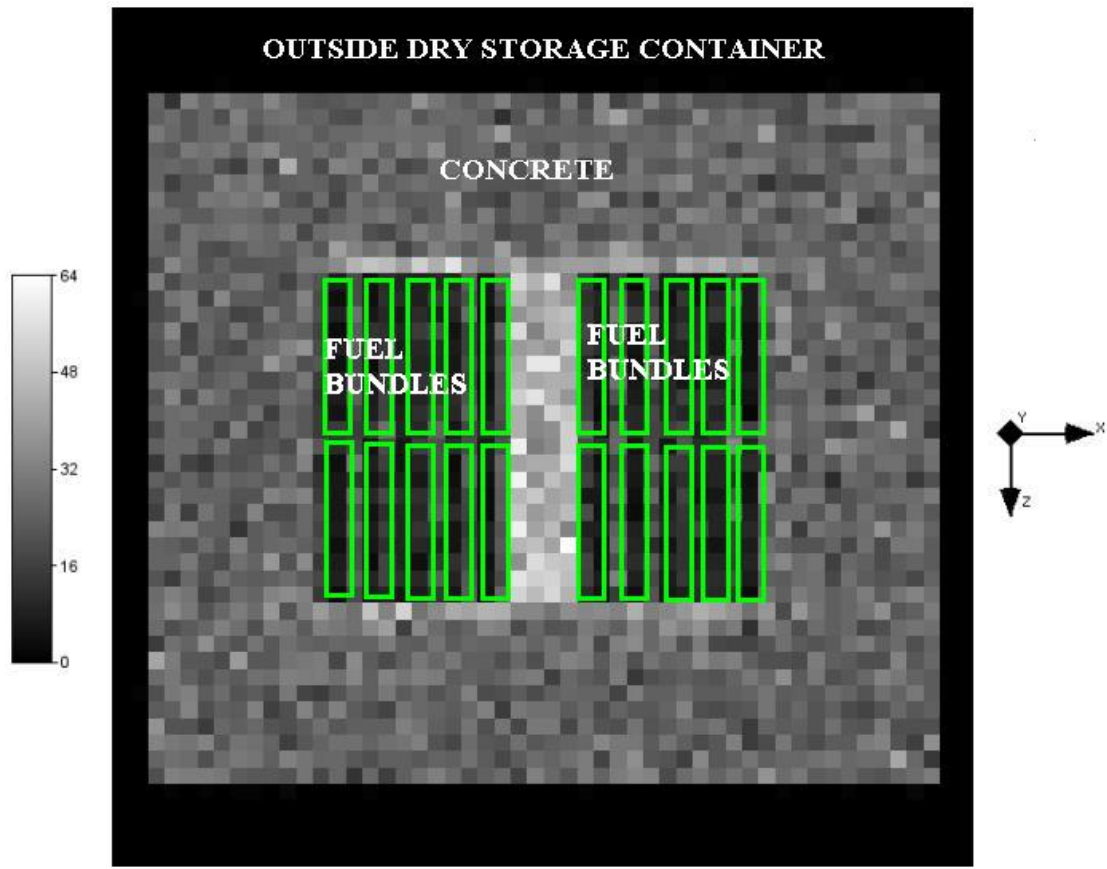

**Figure 5   A muon radiograph of a simulated Dry Storage Container (DSC).**

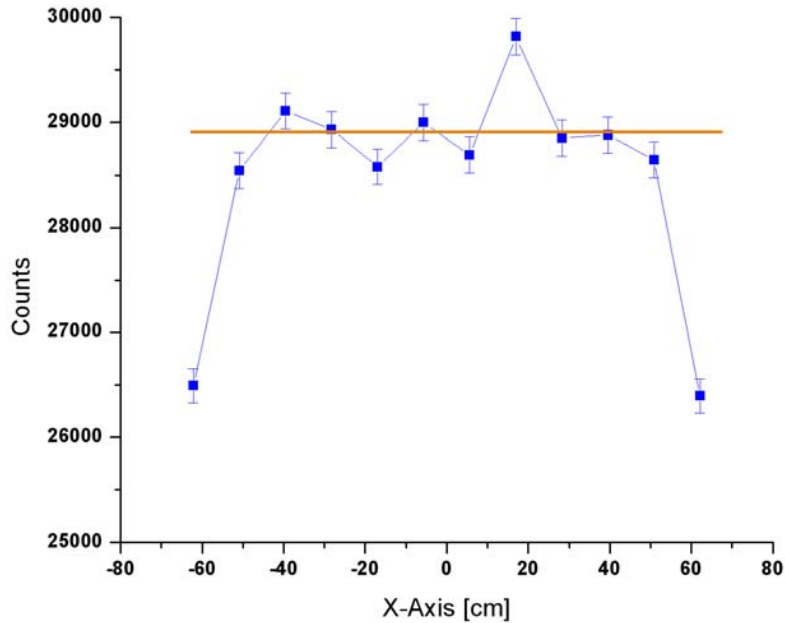

**Figure 6   Simulated muon flux (arbitrary units) projected along one DSC axis.**

**6.      Conclusions**

Owing to the diversity of nuclear waste, it is somewhat unrealistic to look at all possible scenarios in which muon tomography can be used to characterize waste. However, we have shown, through one example, that the method may be useful in identifying clumps of actinides in shielding containers. Future work will investigate the application of the method to the characterization of sludge precipitation in radioactive liquid waste containers and the identification of breach of containment.

This paper outlines the potential application of a muon tomography device for non-security related applications. Muon tomography holds much potential for non-invasive nuclear material accountancy of dry storage containers in that it can potentially improve the accuracy and availability of data on their contents. This can be achieved non-invasively, with the potential for unattended and remotely monitored operation. We have shown that the expected sensitivity to perform material accountancy, in the case of the DSC, greatly exceeds the IAEA detection target for non-proliferation and safeguards.

**7.     Acknowledgements**

The authors wish to acknowledge Bhaskar Sur for suggesting the use of muon tomography as a means to characterize nuclear waste and also for many useful discussions. This work was supported, in part, by the Chemical Biological Radiological and Nuclear Research and Technology Initiative (CRTI) under project CRTI 08-0214RD.